\documentclass[aps,prl,twocolumn,superscriptaddress]{revtex4-1}
\usepackage{amsmath}
\usepackage{graphicx}
\usepackage[outdir=./images/]{epstopdf}
\usepackage[usenames,dvipsnames,svgnames,table]{xcolor}
\usepackage[utf8]{inputenc}
\usepackage{hyperref}
\usepackage{braket}
\usepackage{pbox}
\usepackage{footmisc}
\usepackage[english]{babel}

\begin{document}

	\title{Simulations of Argon Plasma Decay in a Thermionic Converter}

	\author{R. E. Groenewald}
	\affiliation{Modern Electron Inc., Bothell, WA 98011, USA}

	\author{S. Clark}
	\affiliation{Modern Electron Inc., Bothell, WA 98011, USA}

	\author{A. Kannan}
	\affiliation{Modern Electron Inc., Bothell, WA 98011, USA}

	\author{P. Scherpelz}
	\affiliation{Modern Electron Inc., Bothell, WA 98011, USA}

	\date{\today}

	\begin{abstract}

		The dynamics of an argon plasma in the gap of a thermionic diode is investigated using particle-in-cell (PIC) simulations. The time-averaged diode current, as a function of the relative electrical potential between the electrodes, is studied while the plasma density depletes due to recombination on the electrode surfaces. Simulations were performed in both 1D and 2D and significant differences were observed in the plasma decay between the two cases. Specifically, in 2D it was found that the electrostatic potential gradually changes as the plasma decays, while in 1D fluctuations in the plasma led to large potential fluctuations which changed the plasma decay characteristics relative to the 2D case. This creates significant differences in the time-averaged diode current. Furthermore, it was found that the maximum time-averaged current is collected when the diode voltage is set to the flat-band condition, where the cathode and anode vacuum biases are equal. This suggests a novel technique of measuring the difference in work functions between the cathode and anode in a thermionic converter.

	\end{abstract}

	\maketitle

	\subsection{Introduction}

    Thermionic energy converters (TECs) are devices that directly convert heat into electrical energy \cite{hatsopoulosThermionicEnergyConversion1973, goThermionicEnergyConversion2017}. The lack of moving parts and scalability (core conversion efficiency is independent of system size) of the technology makes this type of converter appealing in a wide range of applications \cite{7463074}. Furthermore, thermionics are agnostic to the source of heat used, further widening its potential for impact. Examples of heat sources include solar \cite{clarkSolarThermionicTest2006}, thermonuclear \cite{gyftopoulosThermionicNuclearReactors1963} and natural gas \cite{patent:20200294779}. In its simplest form, a thermionic diode consists of two electrodes physically separated by some gap distance (referred to as the interelectrode gap). One electrode (the cathode) is heated to a temperature at which thermionic emission of electrons occur at a desired current density. The emitted current density is given by the Richardson equation, which relates the thermionically emitted current density to the electrode temperature and work function \cite{crowellRichardsonConstantThermionic1965}. The other electrode (the anode) absorbs some of the thermionically emitted electrons. If the two electrodes are externally connected across a load, this process drives an electrical current through the circuit. Depending on the work functions of the two electrodes and the applied bias from the external circuit, the system can be either power producing or power consuming.
	\begin{figure}[b]
		\includegraphics[width=0.8\columnwidth]{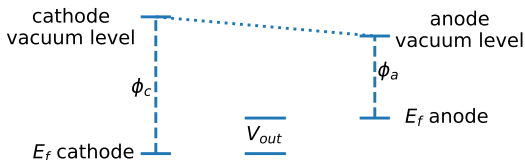}
		\caption{Electron motive diagram to show relation of different physical parameters of interest in a thermionic diode. $E_f$ indicates Fermi levels, $\phi_c$ is the cathode work function, $\phi_a$ the anode work function and $V_{out}$ the output voltage of the device.}
    \label{fig:energy_diagram}
	\end{figure}
Figure \ref{fig:energy_diagram} shows a typical electron motive diagram of a thermionic diode. The electron motive diagram plots the negative of the electrical potential ($\phi = -eV$, where $e$ is the electronic charge and $V$ is electrostatic potential). The device produces power whenever the anode Fermi level is at a lower electrical potential than the cathode's i.e. $V_{out} > 0$. The diagram also shows that as long as $V_{out} + \phi_a < \phi_c$, electrons are accelerated towards the anode, called the accelerating regime. Conversely, when $V_{out} + \phi_a > \phi_c$, electrons are decelerated as they move towards the anode, termed the retarding regime. The middle point, when $V_{out} + \phi_a = \phi_c$, is called the flat-band condition. This corresponds to the case where $V_{out}$ equals the contact potential difference, $\phi_c - \phi_a$.

\begin{figure}
		\includegraphics[width=0.55\columnwidth]{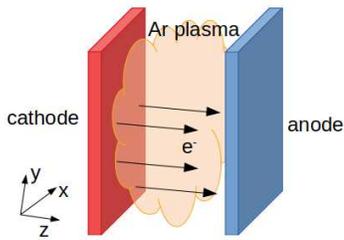}
		\caption{Schematic of a plasma-based thermionic diode. The diode consists of two parallel plates, the cathode (left) and anode (right). Arrows indicate the direction of motion of thermionically emitted electrons as they travel from the cathode to the anode. An argon plasma is present in the interelectrode gap to neutralize the space charge formed by the emitted electrons, and thereby facilitates the flow of current.}
    \label{fig:tec_schematic}
	\end{figure}

    In reality, for a vacuum thermionic diode in which the interelectrode gap is more than a few microns wide, only a small fraction of the emitted electrons make it through the gap to the anode. A space charge barrier forms in front of the cathode that reflects most of the emitted electrons back to the cathode. In such a case the diode current obeys the Child-Langmuir law \cite{childDischargeHotCao1911,langmuirEffectSpaceCharge1913}. Various strategies have been reported to increase the fraction of emitted current that makes it to the anode. Belbachir et al. (2014) \cite{belbachirThermalInvestigationMicrogap2014} used spacers to maintain a sufficiently small interelectrode gap (10 $\mu$m) to avoid the space charge problem. Meir et al. (2013) \cite{meirHighlyefficientThermoelectronicConversion2013} and Wanke et al. (2016) \cite{wankeMagneticfieldfreeThermoelectronicPower2016} used additional electrodes biased positively to reduce the space charge barrier height. Unfortunately, these methods have so far been unable to produce stable, long term operation of a thermionic converter. Another approach, used more successfully in previous TEC development programs, is to use positive ions in the gap to neutralize the space charge barrier \cite{rasorThermionicEnergyConversion1991, baksht}. Cesium plasmas have been heavily employed for this purpose, since the ionization energy of cesium is low \cite{hernqvistAnalysisArcMode1963,wilkinsThermionicConvertersOperating1966}. While these plasmas are effective at mitigating space charge, the energy required to maintain an arc-discharge, defined as the arc-drop, heavily reduces the efficiency of the converter \cite{goThermionicEnergyConversion2017}. This is due to the large neutral scattering cross section of cesium atoms for low energy electrons. As a solution to this problem, using inert gas plasmas have been suggested where the Ramsauer minimum makes these gases mostly transparent to low energy electrons. Using an inert gas plasma requires the plasma ignition be engineered to be very energy efficient. This can be achieved by including highly biased auxiliary electrodes that are optimally placed to produce an inert gas plasma in the gap \cite{oettingerExperimentsEnhancedMode1978, huffman1976high}. Another way is to apply short high voltage pulses across a diode \cite{Zherebtsov, mcveyImprovedPulsedIonization1990}. As electrons accelerate towards the anode they collide with neutral atoms in the gap, causing ionization and eventually striking a plasma in the gap. A schematic of this is shown in Fig.~\ref{fig:tec_schematic}. When the plasma producing pulse is turned off, the plasma starts to slowly decay to the electrodes (the dynamics of this decay are discussed in detail later). During this time, a large portion of the space charge cloud from thermionically emitted electrons is neutralized and the diode current remains high. Once the plasma density has decayed to such a level where it can no longer support high diode current, the plasma ignition pulse is repeated to start the process over again. By keeping a very low duty cycle of on-to-off phases of the ignition pulse, high time-averaged diode current can be sustained with relatively little energy spent to repeatedly strike the plasma \cite{rasorassociatesinc.sunnyvalecalif.usaAdvancedThermionicEnergy1975}.

    While pulsed argon plasma thermionics have been studied concerning the energy required to repeatedly strike the plasma, little is known about the impact of output voltage on the decay dynamics of the plasma, or phrased differently, how the relative bias between cathode and anode affects the average diode current. Understanding the IV-curve of a pulsed plasma converter is important since often this is the only diagnostic available which researchers have to deduce several system parameters such as cathode and anode work functions, plasma density, gap, etc. In the following, this question is explored using particle-in-cell (PIC) simulations. The decay of an argon plasma in the gap of a thermionic converter as a function of time was studied with different biases applied to the anode relative to the cathode. Specifics of the simulation setup are discussed in the Methods section. It was found that the plasma lifetime is strongly dependent on the anode bias. Thereby a characteristic of the IV-curve was identified that allows extraction of the difference in electrode work functions. Furthermore, simulations were performed in both 1D and 2D, and it was found that the decay characteristics of the plasma are strongly dependent on the dimensionality of the system. The Results section describes the differences in the plasma decay under different anode biases as well as the differences that arise from dimensionality. In the Discussion section the plasma sheaths are studied and the origin of the dimensional dependence is explored.

	\subsection{Theoretical considerations}

    A well documented characteristic of pulsed inert-gas plasma thermionic converters is that once the plasma producing pulse is turned off, the plasma sheath in front of the cathode rapidly inverts from an ion-accelerating to an ion-retaining (electron-rich) sheath \cite{rasorassociatesinc.sunnyvalecalif.usaAdvancedThermionicEnergy1975, campanellImprovedUnderstandingHot2017}. This is a desirable situation, since if instead the plasma producing pulse created a dense enough plasma to completely mitigate the thermionic space charge barrier for an extended amount of time, it means for some number of produced ions their decay doesn't affect the diode current. Therefore, the energy used to create those ions was wasted. It consequently would be more energy efficient to pulse for a shorter amount of time and instead re-pulse more frequently so that each produced ion has a maximal impact on the power producing diode current. For this reason, we can assume that during the power producing phase, the cathode sheath will always be ion-retaining. According to McVey (1990) \cite{eindhovenConference1989} the electron and ion currents can be described by the following equations:
\begin{subequations}\label{eq:constraints}
    \begin{align}
        \label{eq:elec_current_cathode}
        J_e &= \frac{2}{3}\left(J_R\exp\left[\frac{eV_c}{kT_c}\right] - J_{re} \right)\\
        \label{eq:ion_current_cathode}
        J_i &= \left(J_{i} / 2 - J_{ri} \right) \exp\left[\frac{eV_c}{kT_c}\right]
    \end{align}
\end{subequations}
where $J_R$ is the Richardson current emitted from the cathode, $V_c$ is the height of the sheath in front of the cathode, $T_c$ is the cathode temperature, and $J_{r\{i,e\}}$ are the random ion/electron currents given by,
\begin{equation}
    J_{r\{i,e\}} = \frac{en}{4}\sqrt{\frac{8k_BT_{\{i,e\}}}{\pi M_{\{i,e\}}}},
\label{eq:random_currents}
\end{equation}
where $e$ is the electron charge, $n$ the plasma density, $T$ the species temperature in the bulk plasma, $k_B$ Boltzmann's constant, and $M$ the particle mass. Notice that the same barrier, $V_c$, that hinders thermionically emitted electrons from entering the plasma also retains ions in the bulk plasma. This clearly indicates that achieving higher diode currents also leads to faster ion decay.

The direction of the anode sheath depends on the applied bias. Warner and Hansen (1967) \cite{warnerTransportEffectsElectron1967} noted that if the anode vacuum potential is sufficiently low (compared to the cathode's), the anode sheath will be electron-retaining (ion-accelerating). In this configuration the ion current at the anode is simply given by $J_i = 2J_{ri}$. Seeing as the lack of an anode barrier doesn't increase the diode current, this configuration only serves to deplete the ion density, leading to shorter plasma lifetimes and consequently lower time-averaged currents. Clearly, a more favorable configuration is to also have an ion-retaining sheath on the anode side, which similarly as before gives,
\begin{subequations}\label{eq:constraints}
    \begin{align}
        \label{eq:elec_current_anode}
        J_e &= 2J_{re}\\
        \label{eq:ion_current_anode}
        J_i &= \left(J_{i} / 2 + J_{ri} \right) \exp\left[\frac{eV_a}{kT_a}\right].
    \end{align}
\end{subequations}
If we assume a perfectly neutral and uniform plasma without sheaths, an electric field will exist in the gap with $E = (V_c - V_a)/d$, where $d$ is the interelectrode gap distance. This field vanishes when $V_c = V_a$ or equivalently (see Fig.~9 of Ref.~\cite{rasorThermionicEnergyConversion1991}) when $V_{out} = \phi_c - \phi_a$. Although this argument serves to form intuition, in a real system plasma sheaths will be present and their heights will be affected by the energy distribution of the plasma particles. Seeing as the sheaths determine the plasma decay rate, correctly calculating their properties are vital to obtaining an accurate picture of the plasma decay. For this reason PIC simulations were used since the first principle nature of these calculations provide the required accuracy in describing the plasma sheaths.
The PIC simulations discussed in the following sections show that the critical point that leads to the highest time-averaged current corresponds to the flat-band condition.

    The approximate plasma decay constant at flat-band was derived by Rasor (1991) \cite{rasorThermionicEnergyConversion1991}, as
\begin{equation}
    \tau = 2\tau_i \frac{T_i}{T_c}\left[ \frac{2J_s/J_0}{1 + \frac{3}{8}\frac{d}{\lambda}} \right]^{T_c/T_i},
\label{eq:rasor_tau}
\end{equation}
where $\tau_i$ is the ion crossing time, $d$ is the gap distance, $\lambda$ the electron mean free path, $J_s$ the Richardson saturation current density emitted from the cathode, $J_0$ the initial diode current density, $T_i$ the average ion temperature in the gap and $T_c$ the cathode temperature. The ion crossing time, $\tau_i = d/\bar{v}$ can be estimated by noting that the ion current is given by $J_i = ne\bar{v}$, where $n$ is the plasma density and $\bar{v}$ is the drift velocity of the ions (Eq.~\ref{eq:random_currents}), giving
\begin{equation}
    \tau_i = \frac{d}{\bar{v}} = 4d\sqrt{\frac{\pi M_i}{8k_BT_i}}.
\label{eq:tau_i}
\end{equation}
Using typical values of the diode parameters $d = 0.5$ mm, $T_c = 1100$ $^\circ$C, $T_i = 700$ $^\circ$C, $d/\lambda \approx 2$ and $J_s/J_0 \approx 2$ the pulse repetition period is found to be $4\tau \approx 50$ $\mu$s. For this reason the diode current density was averaged over a 50 $\mu$s time interval of the plasma decay in the simulations discussed in this article.

	\subsection{Methods}

    The simulations discussed in this article were performed with an adapted version of the PIC code \textit{Warp} \cite{warp,friedmanThreeDimensionalParticle1992,groteWARPCodeModeling2005}. Specialized electrostatic field solvers were written to directly solve Poisson's equation for the goemetry of problems studied here. The 1D solver uses Gaussian elimination to efficiently solve the 1D Poisson equation\cite{birdsall1985plasma} while the 2D solver uses superLU\cite{li05} to decompose the finite difference matrix and quickly solve the linear system. This was found to be much faster than the multigrid solver implemented in \textit{Warp}. MCC handling of particle collisions\cite{birdsallParticleincellChargedparticleSimulations1991} was added and extensively benchmarked against the results of Turner et al. (2013)\cite{turnerSimulationBenchmarksLowpressure2013} to ensure accuracy (see Supplemental Material). The argon cross-sections parameterized by Phelps (1994)\cite{phelpsApplicationScatteringCross1994} were used through the LXCat Phelps database, including elastic scattering processes as well as ion-neutral charge exchange. All simulations were done with spatial resolution of 0.7 $\mu$m. This value was chosen as it is at least 30$\%$ less than the Debye length of the densest plasma simulated, which was 1.03 $\mu$m. The PIC timestep was chosen as the maximum value such that the CFL condition is still satisfied, assuming a maximum electron energy of 5 eV (much higher than the average energy simulated). This resulted in a timestep of $6.15\times 10^{-13}$ s. The 1D simulations injected 10 macroparticles per timestep and assumed an emission area of 1 m$^2$. The macro-particle weighting was then determined by the Richardson saturation current for the temperature conditions simulated. The 2D simulations used 24 cells in the x-direction from which 1 macro-particle was emitted per cell every timestep. The $y$-length was taken as 1 m for the calculation of the emission area and consequently the macro-particle weight. In 1D (2D) the initial plasma density was simulated with 1000 (100) macro-particles per cell. Particle splitting was implemented to continuously ensure sufficiently high particle count per cell during the plasma decay. The particle splitting algorithm ran every 500 simulation steps. It identified cells in which the particle count was less than 100 (10) for 1D (2D) and doubled the particle count in those cells by cloning all its particles while halving the weight. A convergence study was done to ensure simulation parameters were appropriately chosen as shown in the Supplemental Material. The simulations were performed by introducing a quasi-neutral plasma of a specified density between two parallel plates, as in the schematic in Fig.~\ref{fig:tec_schematic}. The parallel plates are modelled as perfect conductors, thereby creating Dirichlet boundary conditions for the ends of the $\hat{z}$ domain and charges are absorbed when entering the conductor domains. During 2D simulations, periodic boundary conditions are used for the $\hat{x}$ domain. The conductor plate on the left of the computational domain will be referred to as the `cathode', and is modelled as a thermionic emitter i.e. electrons are emitted from the face of the conductor. The emitted electrons have velocities sampled from the thermionic emission distribution derived in the Supplemental Material. The conductor plate on the right of the computational domain will be referred to as the `anode'. In all simulations the cathode vacuum potential is used as the zero potential reference while the anode's vacuum potential is varied in order to study the impact of changing the output voltage of the thermionic diode. The simulations are seeded with a neutral argon plasma of specified peak density. The seeded plasma density follows a sine-distribution that peaks in the middle of the gap. Simulations of plasma ignition indicate that this is close to the expected plasma density profile (see Supplemental Material for further details), confirming the same result from Ref.~\cite{lawlessAnalyticalModelThermionic1986}. The seed ions are assumed to be at the neutral gas temperature (for simplicity taken as the average of the cathode and anode temperatures) while the seed electrons are injected with a temperature equal to the cathode temperature, an assumption commonly made in modelling the electrons in inert-gas plasma thermionic converters, see Ref.~\cite{rasorThermionicEnergyConversion1991} for example. This approach has two main shortcomings, namely, in a real pulsed mode TEC the plasma particles would have non-zero drift velocity due to the ignition pulse and the particle density at the electrode surfaces are not as low as predicted by the sine function. Nonetheless this approach is used to avoid the computationally intractable problem of simulating multiple pulse periods.

    The PIC simulations are evolved up to 50 $\mu$s during which the current through the diode is continuously tracked, along with several other quantities such as the spatially resolved plasma density and electrostatic potential. Simulations were performed varying several aspects of the system including the spacing between the electrodes, the density of the initial plasma, the current density emitted from the cathode, and the density of the background neutral gas.

	\subsection{Results}

	\textbf{2D Simulations.}
	\begin{figure}
		\includegraphics[width=1.0\columnwidth]{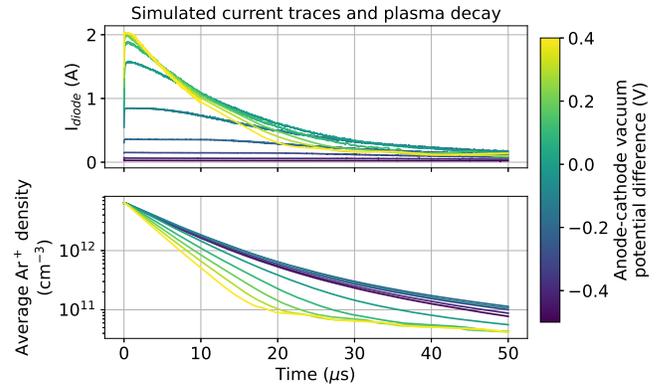}
		\caption{Simulation results for a system with a 500 $\mu$m interelectrode gap, 10 Torr background argon, initial plasma density of $10.2 \times 10^{12}$ cm$^{-3}$ and 2.2 A/cm$^2$ thermionic current emitted from the cathode ($T_c=1100$ $^\circ$C and $\phi_c = 2.1$ eV). The diode current as a function of time is shown in the top panel for different anode potentials. The average plasma density in the gap is shown in the bottom panel.\label{fig:500um_timeseries}}
	\end{figure}
	\begin{figure}[b]
		\includegraphics[width=1.0\columnwidth]{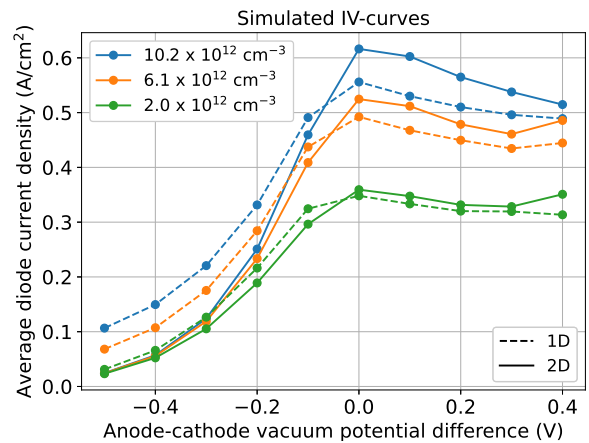}
		\caption{Time-averaged diode current for the output voltage cases shown in Fig.~\ref{fig:500um_timeseries} as well as cases with different initial plasma densities. In all three cases the time-averaged current peaks at the flat-band condition. Results for both 1D and 2D simulations are shown, highlighting the differences in simulation results.\label{fig:500um_iv_curve}}
	\end{figure}
    The first set of results, shown in Fig.~\ref{fig:500um_timeseries}, is for a system where the interelectrode gap was set to 500 $\mu$m, the initial plasma density was set to $10.2 \times 10^{12}$ cm$^{-3}$, and the cathode emission current density was set to 2.2 A/cm$^2$. The simulation results show that as the diode output voltage is increased, moving into the retarding regime, the diode current is suppressed. As the diode output voltage is decreased, moving into the accelerating regime, the diode current at the start of the simulation is increased, but the plasma lifetime is significantly decreased. This leads to a decrease in the diode current as the plasma density diminishes.
The highest time-averaged diode current is seen when the diode is in the flat-band configuration, as shown in Fig.~\ref{fig:500um_iv_curve}. The same simulations were done with different initial plasma densities for which the time-averaged diode current is also shown in Fig.~\ref{fig:500um_iv_curve}. It was found in all simulated cases that the time-averaged current peaks at flat-band. This same result was also seen with simulations of different gap values (250 $\mu$m, 1 mm), different background pressures (15 Torr, 25 Torr) and other emission current densities (8.5 A/cm$^2$, 3.9 A/cm$^2$).\\

	\textbf{1D Simulations.}
    The computational benefit of being able to do 1D simulations that accurately describe systems with translational invariance is clear. Unfortunately, it was found that the specific simulations discussed in this paper do not have the same results when performed in 1D as in 2D.

	\begin{figure}[b]
		\includegraphics[width=1.0\columnwidth]{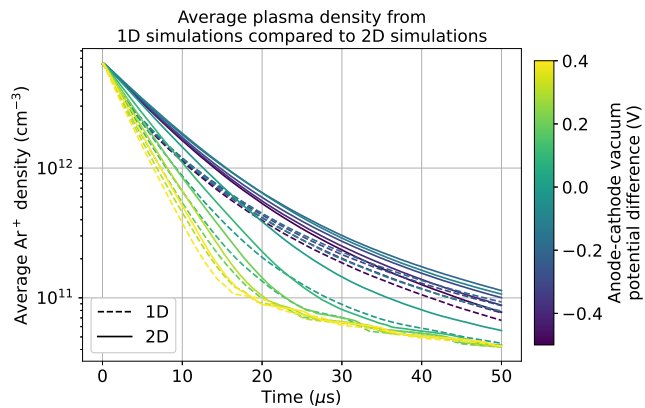}
		\caption{Average plasma density as a function of time for different output voltages showing results from 1D (dashed) and 2D (solid) simulations. Note the differences in the plasma density between the two cases, which is the cause for the differences in time-averaged collected current. The simulation parameters were as follows: 500 $\mu$m gap, 10 Torr background argon, initial plasma density of $2.2 \times 10^{12}$ cm$^{-3}$, $T_c=1100$ $^\circ$C and $\phi_c=2.1$ eV.}
    \label{fig:1d_vs_2d}
	\end{figure}

    As shown in Fig.~\ref{fig:1d_vs_2d}, it was found that the plasma decay simulated in 2D did not match results from performing the same simulation in 1D. Specifically, 1D simulations consistently showed faster decay of the plasma density at the beginning of the simulation compared to the 2D simulations, which then slowed down significantly as the simulation progressed. This leads to lower average current densities in the accelerating regime where the plasma density in the early times of the simulations matters most and higher average current densities in the retarding regime where the slower decay late in the simulation matters more, when compared to the 2D results (see Fig. \ref{fig:500um_iv_curve}).

%	\begin{figure}
%		\includegraphics[width=1.0\columnwidth]{./images/iv_curve_1100_1d_vs_2d}
%		\caption{The result from the 2D simulation shown in Fig.~\ref{fig:500um_iv_curve} with the results of the same simulation parameters from a 1D simulation overlaid. Notice that the 1D simulations drastically overestimates the average diode current away from flat-band but underestimates it close to flat-band.}
%    \label{fig:1d_vs_2d_iv_curve}
%	\end{figure}

	\begin{figure}
		\includegraphics[width=1.0\columnwidth]{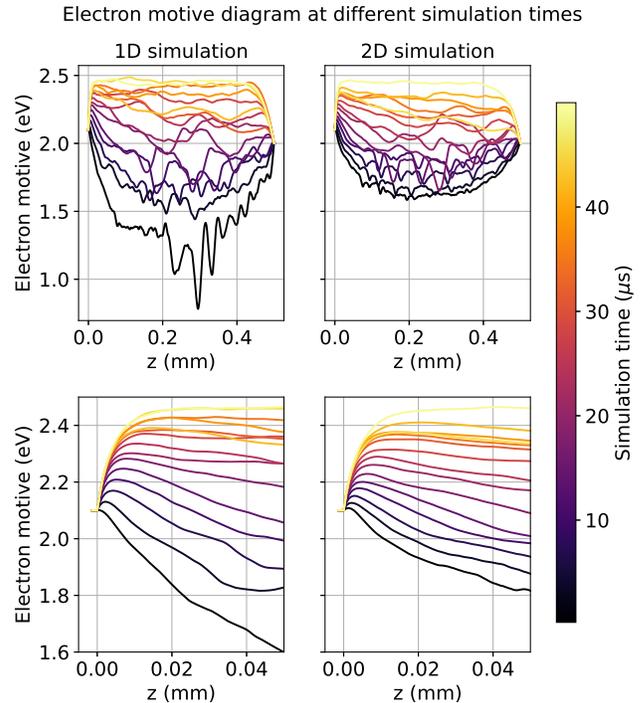}
		\caption{Time evolution of the electron motive for both 1D and 2D simulations, (top) over the full domain and (middle) zoomed in near the cathode. The simulations are started with the same plasma density ($10.2 \times 10^{12}$ cm$^{-3}$), electron temperature, and ion temperature. The case shown is where the anode vacuum level is biased 0.1 V relative to the cathode's.}
    \label{fig:1d_vs_2d_electron_motive}
	\end{figure}

	\begin{figure}
        \includegraphics[width=1.0\columnwidth]{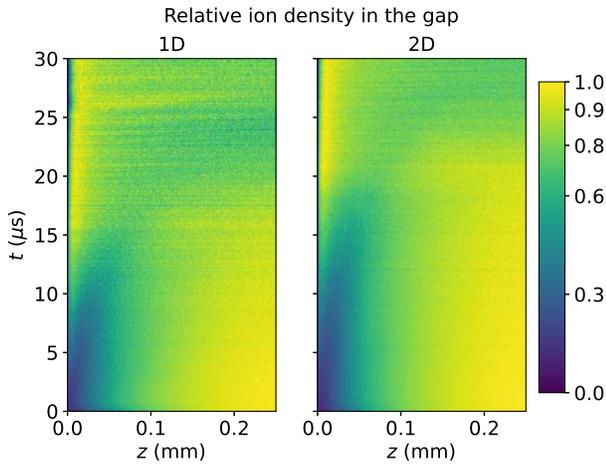}
		\caption{Time evolution of the ion density distribution is shown for a region in front of the cathode for the same simulation case as shown in Fig. \ref{fig:1d_vs_2d_electron_motive}.}
    \label{fig:1d_vs_2d_ion_density}
	\end{figure}

	\begin{figure}[b]
		\includegraphics[width=1.0\columnwidth]{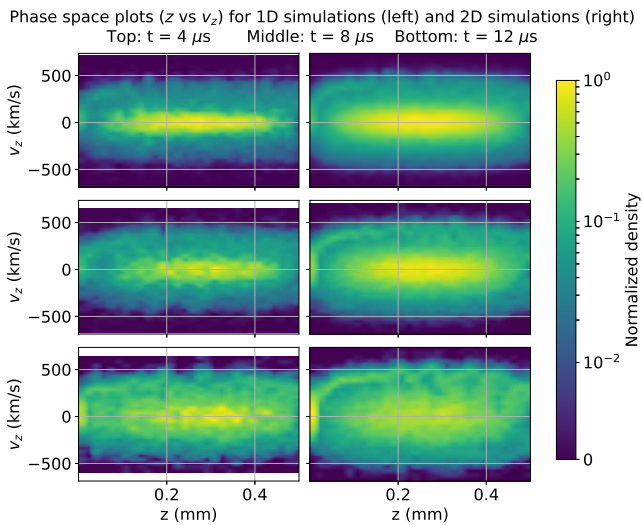}
		\caption{Phase space plots of $z$ versus $v_z$ for different times of the same simulation shown in Fig.~\ref{fig:1d_vs_2d_electron_motive}.}
    \label{fig:phase_space}
	\end{figure}

    The difference in plasma decay characteristics between the 1D and 2D simulations are manifested in the electron motive evolution, as shown in Fig.~\ref{fig:1d_vs_2d_electron_motive}. In 2D, the plasma screening is well captured leading to relatively smooth electrostatic potential profiles throughout the plasma decay. The time-averaged RMS deviation between the motive at $x=0$ and the motive averaged over the x-domain is only $61$ meV, showing that small charge inhomogeneities are well screened by the surrounding plasma. In 1D, however, the reduced dimensionality of the simulations is unable to capture the plasma screening sufficiently leading to abrupt changes in the electrostatic potential as small regions of charge separation form. At and above flat-band, these potential variations lead to a higher average barrier for beam electrons (see Fig.~\ref{fig:1d_vs_2d_electron_motive}) resulting in an under-prediction of the diode current compared to the 2D case. The bottom panel in Fig.~\ref{fig:1d_vs_2d_electron_motive} shows the evolution of the cathode sheath. Initially the potential from the cathode is much steeper in 1D than in 2D which explains why the initial plasma decay is faster in 1D than in 2D (ions are accelerated into the cathode with greater force in 1D). In both 1D and 2D simulations the sheath starts out at the same potential as the cathode, but quickly decreases in energy as the plasma decays and an ion-retaining sheath forms. The so-called "inverse"-mode described by Campanell (2018)\cite{PhysRevE.97.043207} is seen in both cases towards the end of the simulation. In Ref.~\cite{campanellImprovedUnderstandingHot2017} Campanell and Umansky argue that this state is formed due to charge-exchange collisions between ions and neutrals in the cathode sheath trapping ions and leading to a broadening of that sheath towards the anode. Evidence of this mechanishm is seen in Fig. \ref{fig:1d_vs_2d_ion_density} where it is clear that ion density shifts from the center of the gap to the cathode sheath. It is also notable that the 1D simulation shows a much earlier accumulation of ions in that region which explains why the average diode current is lower for the 1D simulation in the accelerating regime.\\
A cut of the electron phase space ($z$ vs $v_z$) is shown in Fig.~\ref{fig:phase_space} at different times. The phase space plots show a beam instability in both 1D and 2D. However, in 2D the wavelength of the beam instability increases as time progresses, indicating a damping of its growth (not seen in 1D). This damping is due to transverse scattering of the electrons off electrostatic waves, something that cannot happen in 1D since there $\vec{E} = E\hat{z}$. Evidence of this is that at $t = 12\ \mu$s, the transverse temperature, $T_\perp = \frac{m_e}{2k_B}\left< v_\perp^2 \right>$, is $2.1 \times T_{cathode}$ in 2D but only $1.5 \times T_{cathode}$ in 1D. Greiner et al. (1995)\cite{greinerNonlinearDynamicalBehavior1995} reported a similar formation of vortices in the electron phase space diagram of a thermionic converter. They used 1D PIC simulations to study this instability (a specific electrostatic instability known as a Pierce-Buneman instability) and observed the same undamped behavior in their 1D simulations. However, they specifically studied the plasma dynamics at high anode bias, not close to flat-band. Levko (2015)\cite{levkoInfluenceAnodeProcesses2015} also showed similar beam instabilities in 1D PIC simulations of a nitrogen plasma in a thermionic converter and studied how they are affected by electron reflection from the anode.

	\subsection{Discussion}

\begin{figure}
		\includegraphics[width=1.0\columnwidth]{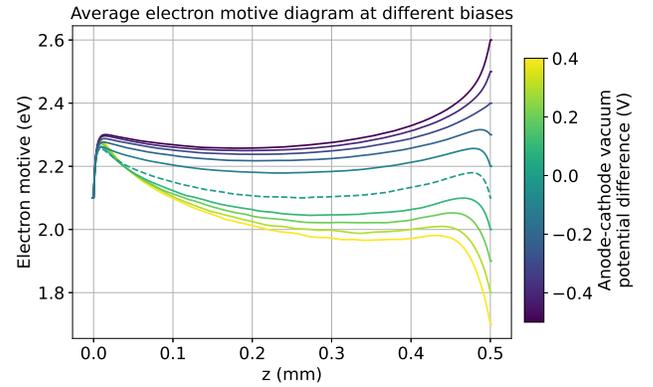}
		\caption{The electron motive averaged over time for different anode bias cases. The flat-band condition is shown with a dashed line to highlight it. The simulation cases shown here are the same as is shown in Fig.~\ref{fig:500um_timeseries}.}
    \label{fig:2d_electron_motive}
	\end{figure}

    The impact of diode output voltage was studied for argon plasma-based thermionic diodes. It was found that the maximum time-averaged current is collected when the diode is in the flat-band configuration. This result can be understood by studying the average electron motive for differently biased cases, as shown in Fig.~\ref{fig:2d_electron_motive}. These results confirm the intuition discussed earlier. At highly negative biases the anode sheath becomes ion-accelerating which understandably leads to low electron current. At highly positive biases the bulk plasma potential is higher than the cathode's vacuum potential. In this condition ions from the bulk plasma easily have enough energy to overcome the ion-retaining sheaths in front of the cathode, causing fast plasma decay. Towards the flat-band condition, the sheaths in front of both electrodes become ion-retaining and single valued, which is the desired condition for slow plasma decay (as discussed earlier). The barrier index (height of cathode sheath relative to the cathode potential) is lowest exactly at flat-band which explains why the maximum time averaged diode current is seen at flat-band.

    This result provides a novel technique to measure the difference in work functions between the cathode and anode in a thermionic converter. In practice it is difficult to determine electrode work functions in operating thermionic diodes. Typically the work functions will be sensitive to temperature, as is the case with dispenser cathodes \cite{levushMigrationEscapeBarium2000} and refractory metal electrodes that rely on cesium-oxide coverage to achieve useful work functions \cite{desplatRecentDevelopmentsOxygenated1997}. Furthermore, evaporation and deposition of electrode material on the opposite electrode greatly alters the electrode work functions. For these reasons measurements have to be done in operating conditions to be reliable. The results discussed here indicate that an \textit{in situ} measurement of work function differences can be done by sweeping the output voltage of an operating pulsed argon plasma diode while recording the time-averaged current. A peak in the time-averaged IV-curve indicates the difference in electrode work functions. This in turn allows one to study more carefully the impact of changing operating conditions by tracking changes to work functions rather than solely the output power of the converter, which is a convoluted measurement of many factors. Experimental work is currently ongoing to test this proposed method.

    It was also observed that simulating the plasma decay, using a PIC approach, resulted in different behavior in 1D than in 2D. Specifically, higher (lower) average currents were seen in the retarding (accelerating) regime in 1D compared to the 2D simulations. Therefore, in its current implementation, this type of simulation cannot be done with high fidelity in 1D. It is believed that the dominant electron thermalization mechanism simulated is scattering off electrostatic waves, which excludes transverse scattering in the 1D case. Adding an anomalous scattering cross section in this case to enable transverse thermalization could recover fidelity in such simulations. This is left to future work. It was also noted that other authors have studied such 1D beam instabilities in similar systems using PIC codes (Ref. \cite{greinerNonlinearDynamicalBehavior1995, levkoInfluenceAnodeProcesses2015}). Studying the dampening of these instabilities in 2D in further detail is also left to future work.

	% \bibliography{bibliography}

	\section{Acknowledgement}

	The authors are grateful to Phil Miller for his assistance in improving the parallel scaling of the internal version of Warp used for the simulations discussed in this article.

    \section{Author contributions}

    All authors contributed to the code development needed for the discussed simulations. RG ran the simulations and wrote the manuscript with input from the other authors.

	\section{Supplemental Material to: Simulations of Argon Plasma Decay in a Thermionic Converter}

    \section{Velocity distribution of thermionically emitted electrons}

    We start by assuming the velocities of electrons that could participate in thermionic emission (electrons close to the top of the conduction band) can be well described by a Maxwellian distribution function,
\begin{equation}
    f_{\vec{v}}(v_x, v_y, v_z) = \left( \frac{1}{\sqrt{2\pi} \sigma}\right)^3\exp\left[ -\frac{v_x^2 + v_y^2 + v_z^2}{2 \sigma^2}\right],
\label{eq:maxwell}
\end{equation}
where $\sigma = \sqrt{\frac{k_B T}{m}}$ is the thermal velocity of an electron.
Let the cathode work function be $\phi$ and assume an idealized step function in the potential between the interior of the cathode and the vacuum. By assumption there is no barrier to escape in the $\hat{x}$ or $\hat{y}$ directions so these velocity components are not perturbed. In the $\hat{z}$ direction, however, only particles with $v_z > 0$ and $\frac{1}{2}mv_z^2 > \phi$ can escape from the cathode. Hence, the electrons that can be thermionically emitted are characterized by
\begin{equation}
    v_z^* > \sqrt{\frac{2\phi}{m}}.
\label{eq:thermionic_vz_limit}
\end{equation}
The cumulative distribution function for the longitudinal velocity of electrons that will be thermionically emitted just prior to emission, is thus given by:
\begin{equation}
    P\left( v_z \leq v_z^* \middle| v_z > \sqrt{\frac{2\phi}{m}}\right) =
    \begin{cases}
        0,      & v_z^* < \sqrt{\frac{2\phi}{m}}\\
        \frac{A}{B},      & v_z^* \geq \sqrt{\frac{2\phi}{m}}
    \end{cases}
\label{eq:cdf_vz}
\end{equation}
where
\begin{subequations}
    \begin{align}
        A = \int_{\sqrt{\frac{2\phi}{m}}}^{v_z^*} f(v_z) dv_z = \frac{1}{2}\left[ \text{erf} \left( \frac{v_z^*}{\sqrt{2}\sigma}\right) -\text{erf} \left( \frac{\sqrt{2\phi/m}}{\sqrt{2}\sigma}\right) \right]\\
        B = \int_{\sqrt{\frac{2\phi}{m}}}^{\infty} f(v_z) dv_z = \frac{1}{2} \left[ 1 - \text{erf} \left( \frac{\sqrt{2\phi/m}}{\sqrt{2}\sigma}\right) \right].
    \end{align}
\end{subequations}
We can now get the distribution function by differentiating the CDF above, giving
\begin{equation}
    f_{inside}(v_z^*) =
    \begin{cases}
        0,      & v_z^* < \sqrt{\frac{2\phi}{m}}\\
        \frac{\frac{1}{\sqrt{2\pi}\sigma} \exp \left( -\frac{v_z^{*2}}{2\sigma^2} \right)}{ \frac{1}{2}\left[ 1 - \text{erf}\left( \frac{\sqrt{2\phi/m}}{\sqrt{2}\sigma} \right) \right]},      & v_z^* \geq \sqrt{\frac{2\phi}{m}}
    \end{cases}
\label{eq:pdf_vz}
\end{equation}
Finally, we compute the velocity distribution function immediately outside the cathode. Conservation of energy requires that the electron kinetic energy be reduced by $\phi$ upon escaping the cathode, so
\begin{equation}
    v_z' = \sqrt{v_z^{*2} - \frac{2\phi}{m}}.
\label{eq:vz_outside}
\end{equation}
We now use the change-of-variables rule for probability distributions to calculate the distribution of $v_z'$ from Eq.~\ref{eq:pdf_vz}, giving,
\begin{equation}
    f_{outside}(v_z') =
    \begin{cases}
        0,      & v_z' < 0\\
        \frac{v_z'}{\sqrt{v_z'^2 + \frac{2\phi}{m}}} \frac{\frac{1}{\sqrt{2\pi}\sigma}\exp\left( \frac{v_z'^2+2\phi/m}{2\sigma^2} \right)}{ \frac{1}{2} \left[ 1 - \text{erf} \left( \frac{\sqrt{2\phi/m}}{\sqrt{2}\sigma}\right) \right] },      & v_z' \geq 0
    \end{cases}
\label{eq:pdf_vz_prime}
\end{equation}
Seeing as there is no barrier in the $\hat{x}$ or $\hat{y}$ directions we have
\begin{subequations}
    \begin{align}
        f_{outside}(v_x') &= f_{inside}(v_x)\\
        f_{outside}(v_y') &= f_{inside}(v_y).
    \end{align}
\end{subequations}

    In the typical case of thermionic converters, cathode work functions are on the order of a few eV while temperatures are on the order of 0.1 eV, thus $k_BT << \phi$. In this regime we can take
$$ v_z'^2 + \frac{2\phi}{m} \approx \frac{2\phi}{m} $$
and Eq.~\ref{eq:pdf_vz_prime} simplifies to
\begin{equation}
    f_{outside}(v_z') =
    \begin{cases}
        0,      & v_z' < 0\\
        \frac{v_z'}{\sigma^2}\exp\left( -\frac{v_z'^2}{2\sigma^2} \right),      & v_z' \geq 0
    \end{cases}
\label{eq:pdf_vz_prime_simple}
\end{equation}
This distribution is used to sample the $\hat{z}$ direction velocities of electrons injected in the PIC simulations discussed in the main paper.

	\section{Code benchmark}

    The MCC implementation, we added to Warp, was benchmarked against the results of Turner et. al. (2013)\cite{turnerSimulationBenchmarksLowpressure2013}. The results of this benchmark is given in Fig.~\ref{fig:turner_benchmark}

	\begin{figure}
		\includegraphics[width=1.0\columnwidth]{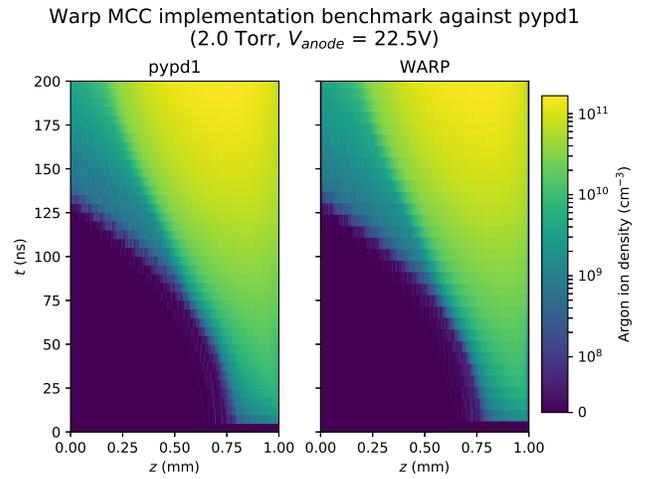}
		\caption{Argon ion density as a function of time and space during plasma ignition as simulated in $\textit{pypd1}$ \cite{verboncoeurObjectorientedElectromagneticPIC1995} compared to the current authors' implementation of the same physics in Warp. Simulated saturation current was 0.14 A/cm$^2$.}
    \label{fig:pypd1_compare1}
	\end{figure}

	\begin{figure}
		\includegraphics[width=1.0\columnwidth]{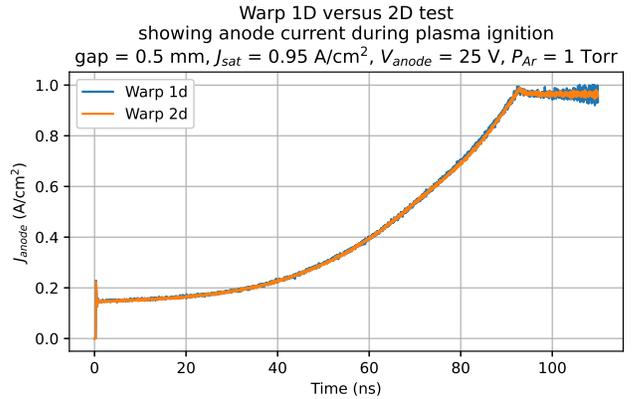}
		\caption{Diode current as a function of time during plasma ignition as simulated in 1D and 2D in $\textit{Warp}$.}
    \label{fig:dim_compare2}
	\end{figure}

	\begin{figure*}
		\includegraphics[width=1.9\columnwidth]{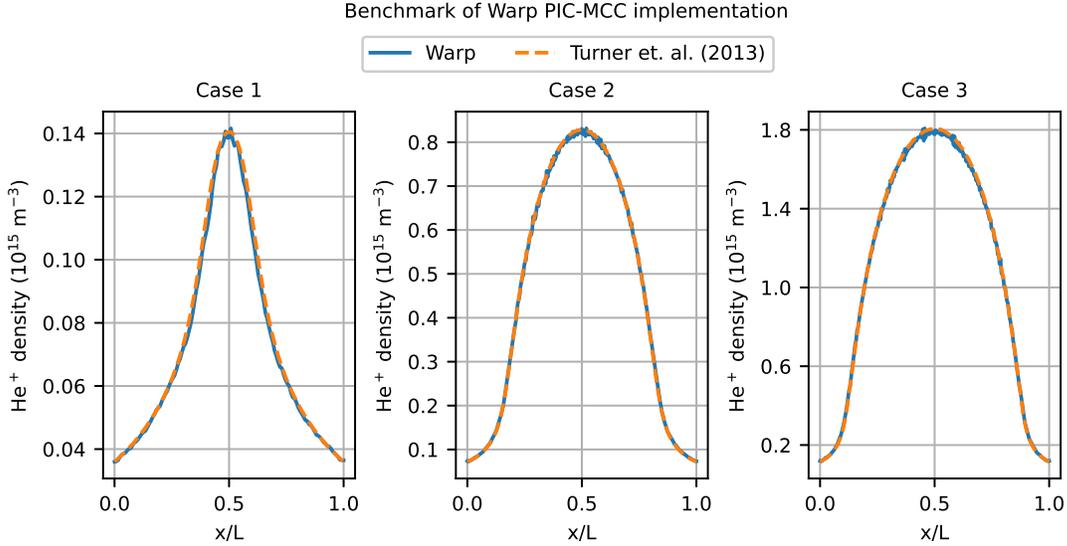}
		\caption{Helium ion density in a capacitively coupled discharge for three cases as presented by Turner et. al. (2013)\cite{turnerSimulationBenchmarksLowpressure2013}, in their work to benchmark PIC-MCC implementations. Results from the current authors' implementation of MCC in $\textit{Warp}$\cite{warp} is overlayed with the literature results to demonstrate accuracy of the implementation.}
    \label{fig:turner_benchmark}
	\end{figure*}

    In order to also test the $\textit{Warp}$ implementation in a more appropriate situation to the simulations performed in this study, results were compared to those from the PIC code $\textit{oopd1}$, through the python wrapper $\textit{pypd1}$ \cite{verboncoeurObjectorientedElectromagneticPIC1995}. The plot in Fig.~\ref{fig:pypd1_compare1} shows the ion density evolution during plasma ignition in a 1 mm diode over 200 ns.

    Finally a test of the 1D versus 2D implementations of the code was done. The plasma ignition in a 500 $\mu$m gap diode with 1 Torr of argon and the anode biased to 25 V was simulated. The current collected on the anode as a function of time as calculated in 1D and 2D in $\textit{Warp}$. Results for this test are shown in Fig.~\ref{fig:dim_compare2}. The test shows that the MCC implementation in the modified $\textit{Warp}$ code behaves the same in 1D as in 2D. This indicates that the differences seen in 1D and 2D as highlighted in the article are due to the dimensional dependence of the problem, not the PIC implementation.

    \section{Plasma density profile after ignition pulse}

\begin{figure}[b]
		\includegraphics[width=1.0\columnwidth]{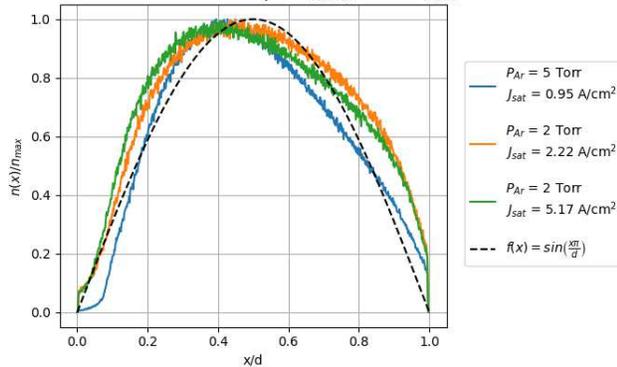}
		\caption{Plasma density after ignition as simulated in Warp, by applying a 25 V bias across a diode of 500 $\mu$m gap for 100 ns. Also shown is a sine function scaled to have a wavelength of $d/2$.}
    \label{fig:plasma_distribution}
	\end{figure}

    The simulations discussed in this article started with seeded plasma densities distributed according to a sine function. This distribution was chosen after simulations of plasma ignition indicated that it is a good approximation for the plasma density formed, as shown in Fig.~\ref{fig:plasma_distribution}. The seed density specified in the labels of figures refers to the plasma density at the maximum point (in the middle of the gap).

    The initial plasma was seeded as follows: Let $X_1, ..., X_N$ be an i.i.d. sample from $U(0, 1)$ (uniformly distributed between 0 and 1), where $N$ is the desired number of seed macro-particles. The $z$ position of each particle was then calculated according to
$$z_i = \frac{D}{\pi}\arccos(1 - 2X_i), $$
where $D$ is the interelectrode gap. The statistical weight of each particle was given by
$$ W = \frac{2}{\pi}\frac{V_{sim}n_{init}}{N w_{e}},$$
where $V_{sim}$ is the volume of the computational domain, $n_{init}$ the desired peak initial density and $w_e$ the macro-particle weight calculated from the specified thermionic emission current density and cathode area combined with the specified NPPC injection rate. In 2D the $x$ position of the seed particles were simply set by another i.i.d. sample from $U(0, x_{max})$. The electron and ion velocities were sampled from Maxwellian distributions at the specified temperatures for each species.

	\section{Convergence Study}

	\begin{figure}
		\includegraphics[width=0.96\columnwidth]{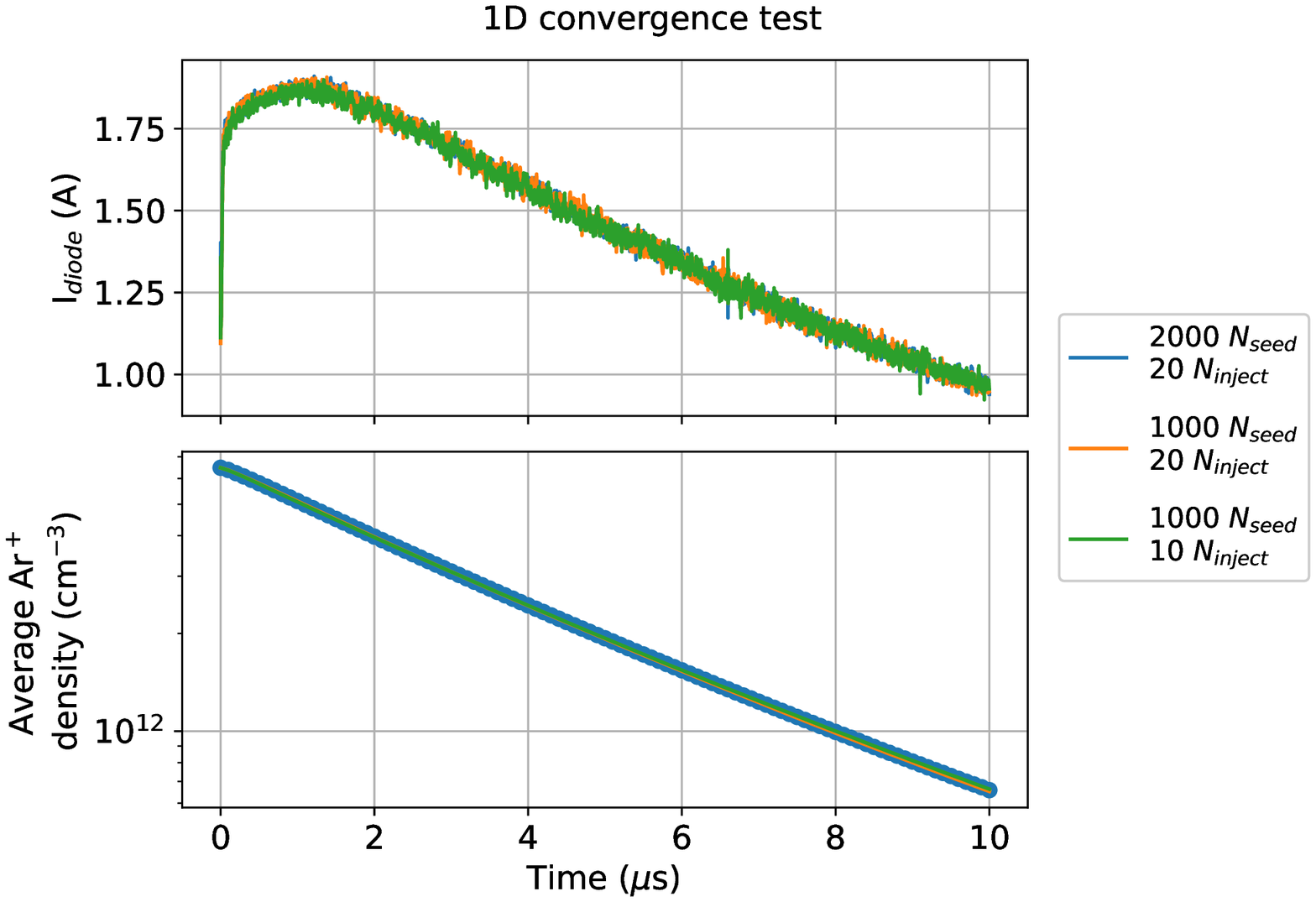}
        \includegraphics[width=0.96\columnwidth]{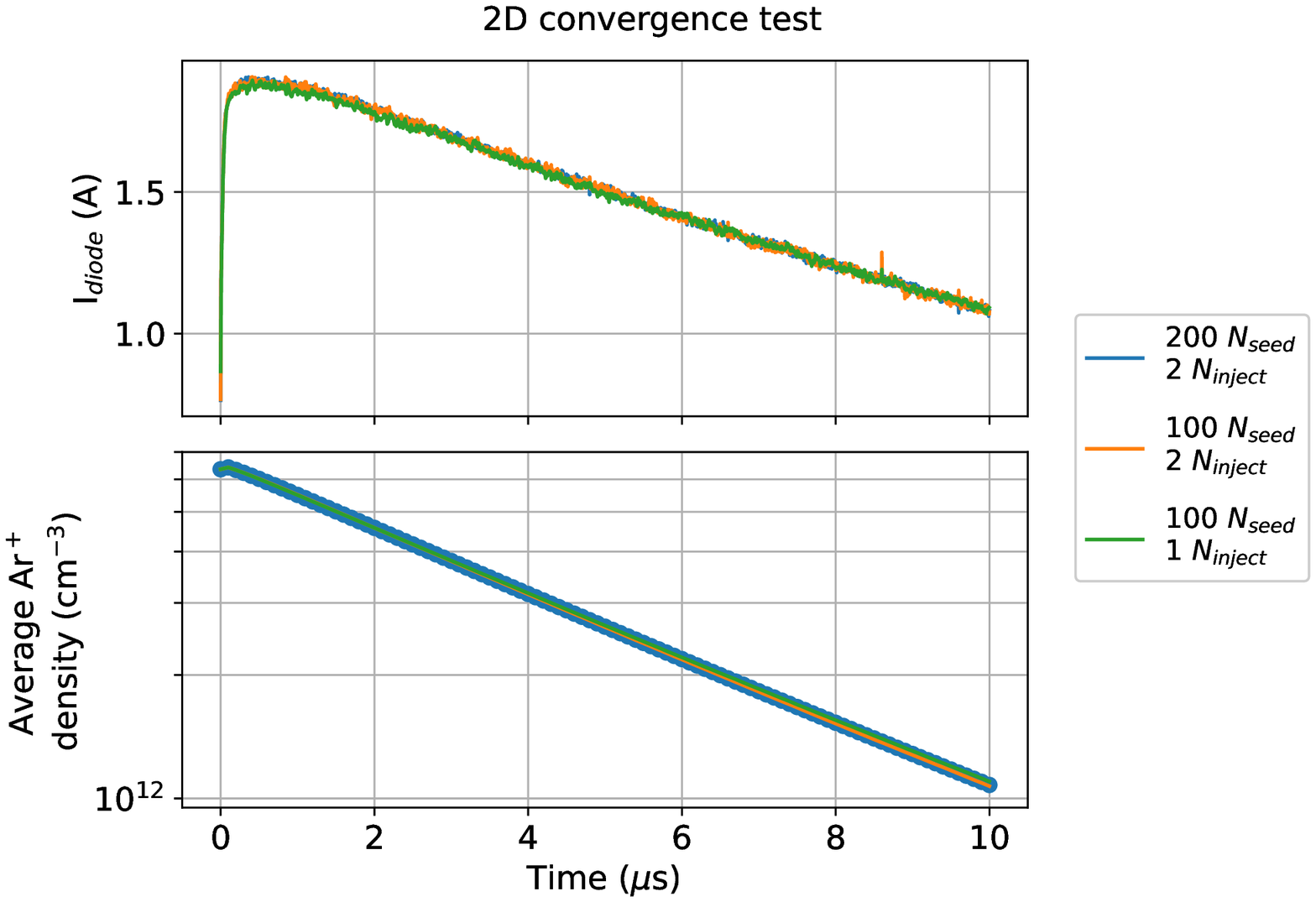}
		\caption{Test of simulation convergence with respect to macro-particle count in 1D (top) and 2D (bottom). The number of macro-particles per cell ($N_{seed}$) used to simulated the seed plasma was varied as well as the number of macro-particles injected per cathode cell per timestep ($N_{inject}$) and the results compared.}
    \label{fig:conv_results}
	\end{figure}

    A study to check whether the initial number of particles used and the number of particles injected per timestep was sufficient to achieve converged results was performed by running a case with a 500 $\mu$m gap, cathode temperature of 1100 $^\circ$C, cathode work function of 2.1 eV, initial peak plasma density of $10.2\times10^{12}$ cm$^{-3}$ and background argon pressure of 10 Torr. The simulation was run up to 10 $\mu$s of simulation time while comparing time traces for the diode current and plasma density. The results for 1D and 2D runs are shown in Fig.~\ref{fig:conv_results}. In the simulations discussed in the main paper 1000 (100) seed particles per cell were used in 1D (2D) while injecting 10 (1) particles per cell per timestep from the cathode. During the convergence study all particle numbers were doubled without any significant changes in the simulation results indicating convergence.

	\bibliography{bibliography}

\begin{thebibliography}{39}%
\makeatletter
\providecommand \@ifxundefined [1]{%
 \@ifx{#1\undefined}
}%
\providecommand \@ifnum [1]{%
 \ifnum #1\expandafter \@firstoftwo
 \else \expandafter \@secondoftwo
 \fi
}%
\providecommand \@ifx [1]{%
 \ifx #1\expandafter \@firstoftwo
 \else \expandafter \@secondoftwo
 \fi
}%
\providecommand \natexlab [1]{#1}%
\providecommand \enquote  [1]{``#1''}%
\providecommand \bibnamefont  [1]{#1}%
\providecommand \bibfnamefont [1]{#1}%
\providecommand \citenamefont [1]{#1}%
\providecommand \href@noop [0]{\@secondoftwo}%
\providecommand \href [0]{\begingroup \@sanitize@url \@href}%
\providecommand \@href[1]{\@@startlink{#1}\@@href}%
\providecommand \@@href[1]{\endgroup#1\@@endlink}%
\providecommand \@sanitize@url [0]{\catcode `\\12\catcode `\$12\catcode
  `\&12\catcode `\#12\catcode `\^12\catcode `\_12\catcode `\%12\relax}%
\providecommand \@@startlink[1]{}%
\providecommand \@@endlink[0]{}%
\providecommand \url  [0]{\begingroup\@sanitize@url \@url }%
\providecommand \@url [1]{\endgroup\@href {#1}{\urlprefix }}%
\providecommand \urlprefix  [0]{URL }%
\providecommand \Eprint [0]{\href }%
\providecommand \doibase [0]{http://dx.doi.org/}%
\providecommand \selectlanguage [0]{\@gobble}%
\providecommand \bibinfo  [0]{\@secondoftwo}%
\providecommand \bibfield  [0]{\@secondoftwo}%
\providecommand \translation [1]{[#1]}%
\providecommand \BibitemOpen [0]{}%
\providecommand \bibitemStop [0]{}%
\providecommand \bibitemNoStop [0]{.\EOS\space}%
\providecommand \EOS [0]{\spacefactor3000\relax}%
\providecommand \BibitemShut  [1]{\csname bibitem#1\endcsname}%
\let\auto@bib@innerbib\@empty
%</preamble>
\bibitem [{\citenamefont {Hatsopoulos}\ and\ \citenamefont
  {Gyftopoulos}(1973)}]{hatsopoulosThermionicEnergyConversion1973}%
  \BibitemOpen
  \bibfield  {author} {\bibinfo {author} {\bibfnamefont {G.~N.}\ \bibnamefont
  {Hatsopoulos}}\ and\ \bibinfo {author} {\bibfnamefont {E.~P.}\ \bibnamefont
  {Gyftopoulos}},\ }\href@noop {} {{\selectlanguage {English}\emph {\bibinfo
  {title} {Thermionic Energy Conversion}}}}\ (\bibinfo  {publisher} {{MIT
  Press}},\ \bibinfo {address} {{Cambridge}},\ \bibinfo {year}
  {1973})\BibitemShut {NoStop}%
\bibitem [{\citenamefont {Go}\ \emph {et~al.}(2017)\citenamefont {Go},
  \citenamefont {Haase}, \citenamefont {George}, \citenamefont {Mannhart},
  \citenamefont {Wanke}, \citenamefont {Nojeh},\ and\ \citenamefont
  {Nemanich}}]{goThermionicEnergyConversion2017}%
  \BibitemOpen
  \bibfield  {author} {\bibinfo {author} {\bibfnamefont {D.~B.}\ \bibnamefont
  {Go}}, \bibinfo {author} {\bibfnamefont {J.~R.}\ \bibnamefont {Haase}},
  \bibinfo {author} {\bibfnamefont {J.}~\bibnamefont {George}}, \bibinfo
  {author} {\bibfnamefont {J.}~\bibnamefont {Mannhart}}, \bibinfo {author}
  {\bibfnamefont {R.}~\bibnamefont {Wanke}}, \bibinfo {author} {\bibfnamefont
  {A.}~\bibnamefont {Nojeh}}, \ and\ \bibinfo {author} {\bibfnamefont
  {R.}~\bibnamefont {Nemanich}},\ }\href {\doibase 10.3389/fmech.2017.00013}
  {\bibfield  {journal} {\bibinfo  {journal} {Frontiers in Mechanical
  Engineering}\ }\textbf {\bibinfo {volume} {3}},\ \bibinfo {pages} {13}
  (\bibinfo {year} {2017})}\BibitemShut {NoStop}%
\bibitem [{\citenamefont {{Abdul Khalid}}\ \emph {et~al.}(2016)\citenamefont
  {{Abdul Khalid}}, \citenamefont {{Leong}},\ and\ \citenamefont
  {{Mohamed}}}]{7463074}%
  \BibitemOpen
  \bibfield  {author} {\bibinfo {author} {\bibfnamefont {K.~A.}\ \bibnamefont
  {{Abdul Khalid}}}, \bibinfo {author} {\bibfnamefont {T.~J.}\ \bibnamefont
  {{Leong}}}, \ and\ \bibinfo {author} {\bibfnamefont {K.}~\bibnamefont
  {{Mohamed}}},\ }\href@noop {} {\bibfield  {journal} {\bibinfo  {journal}
  {IEEE Transactions on Electron Devices}\ }\textbf {\bibinfo {volume} {63}},\
  \bibinfo {pages} {2231} (\bibinfo {year} {2016})}\BibitemShut {NoStop}%
\bibitem [{\citenamefont {Clark}(2006)}]{clarkSolarThermionicTest2006}%
  \BibitemOpen
  \bibfield  {author} {\bibinfo {author} {\bibfnamefont {P.~N.}\ \bibnamefont
  {Clark}},\ }in\ \href {\doibase 10.1063/1.2169240} {{\selectlanguage
  {English}\emph {\bibinfo {booktitle} {{{AIP Conference Proceedings}}}}}},\
  Vol.\ \bibinfo {volume} {813}\ (\bibinfo  {publisher} {{AIP}},\ \bibinfo
  {address} {{Albuquerque, New Mexico (USA)}},\ \bibinfo {year} {2006})\ pp.\
  \bibinfo {pages} {598--606}\BibitemShut {NoStop}%
\bibitem [{\citenamefont {Gyftopoulos}\ and\ \citenamefont
  {Hatsopoulos}(1963)}]{gyftopoulosThermionicNuclearReactors1963}%
  \BibitemOpen
  \bibfield  {author} {\bibinfo {author} {\bibfnamefont {E.~P.}\ \bibnamefont
  {Gyftopoulos}}\ and\ \bibinfo {author} {\bibfnamefont {G.~N.}\ \bibnamefont
  {Hatsopoulos}},\ }\href {\doibase 10.1109/EE.1963.6541300} {\bibfield
  {journal} {\bibinfo  {journal} {Electrical Engineering}\ }\textbf {\bibinfo
  {volume} {82}},\ \bibinfo {pages} {108} (\bibinfo {year} {1963})}\BibitemShut
  {NoStop}%
\bibitem [{\citenamefont {Ashton}\ \emph {et~al.}(2020)\citenamefont {Ashton},
  \citenamefont {Clark}, \citenamefont {Kokonaski}, \citenamefont {Kraemer},
  \citenamefont {Lorr}, \citenamefont {Mankin}, \citenamefont {Menacher},
  \citenamefont {Noble}, \citenamefont {Pan}, \citenamefont {De~Pijper},\ and\
  \citenamefont {Wood}}]{patent:20200294779}%
  \BibitemOpen
  \bibfield  {author} {\bibinfo {author} {\bibfnamefont {J.~B.}\ \bibnamefont
  {Ashton}}, \bibinfo {author} {\bibfnamefont {S.~E.}\ \bibnamefont {Clark}},
  \bibinfo {author} {\bibfnamefont {W.}~\bibnamefont {Kokonaski}}, \bibinfo
  {author} {\bibfnamefont {D.}~\bibnamefont {Kraemer}}, \bibinfo {author}
  {\bibfnamefont {J.~J.}\ \bibnamefont {Lorr}}, \bibinfo {author}
  {\bibfnamefont {M.~N.}\ \bibnamefont {Mankin}}, \bibinfo {author}
  {\bibfnamefont {D.~J.}\ \bibnamefont {Menacher}}, \bibinfo {author}
  {\bibfnamefont {P.~D.}\ \bibnamefont {Noble}}, \bibinfo {author}
  {\bibfnamefont {T.~S.}\ \bibnamefont {Pan}}, \bibinfo {author} {\bibfnamefont
  {A.}~\bibnamefont {De~Pijper}}, \ and\ \bibinfo {author} {\bibfnamefont
  {L.~L.}\ \bibnamefont {Wood}},\ }\href
  {https://www.freepatentsonline.com/y2020/0294779.html} {\enquote {\bibinfo
  {title} {Combined heating and power modules and devices},}\ } (\bibinfo
  {year} {2020})\BibitemShut {NoStop}%
\bibitem [{\citenamefont
  {Crowell}(1965)}]{crowellRichardsonConstantThermionic1965}%
  \BibitemOpen
  \bibfield  {author} {\bibinfo {author} {\bibfnamefont {C.}~\bibnamefont
  {Crowell}},\ }\href {\doibase 10.1016/0038-1101(65)90116-4} {\bibfield
  {journal} {\bibinfo  {journal} {Solid-State Electronics}\ }\textbf {\bibinfo
  {volume} {8}},\ \bibinfo {pages} {395} (\bibinfo {year} {1965})}\BibitemShut
  {NoStop}%
\bibitem [{\citenamefont {Child}(1911)}]{childDischargeHotCao1911}%
  \BibitemOpen
  \bibfield  {author} {\bibinfo {author} {\bibfnamefont {C.~D.}\ \bibnamefont
  {Child}},\ }\href {\doibase 10.1103/PhysRevSeriesI.32.492} {\bibfield
  {journal} {\bibinfo  {journal} {Physical Review (Series I)}\ }\textbf
  {\bibinfo {volume} {32}},\ \bibinfo {pages} {492} (\bibinfo {year}
  {1911})}\BibitemShut {NoStop}%
\bibitem [{\citenamefont {Langmuir}(1913)}]{langmuirEffectSpaceCharge1913}%
  \BibitemOpen
  \bibfield  {author} {\bibinfo {author} {\bibfnamefont {I.}~\bibnamefont
  {Langmuir}},\ }\href {\doibase 10.1103/PhysRev.2.450} {\bibfield  {journal}
  {\bibinfo  {journal} {Physical Review}\ }\textbf {\bibinfo {volume} {2}},\
  \bibinfo {pages} {450} (\bibinfo {year} {1913})}\BibitemShut {NoStop}%
\bibitem [{\citenamefont {Belbachir}\ \emph {et~al.}(2014)\citenamefont
  {Belbachir}, \citenamefont {An},\ and\ \citenamefont
  {Ono}}]{belbachirThermalInvestigationMicrogap2014}%
  \BibitemOpen
  \bibfield  {author} {\bibinfo {author} {\bibfnamefont {R.~Y.}\ \bibnamefont
  {Belbachir}}, \bibinfo {author} {\bibfnamefont {Z.}~\bibnamefont {An}}, \
  and\ \bibinfo {author} {\bibfnamefont {T.}~\bibnamefont {Ono}},\ }\href
  {\doibase 10.1088/0960-1317/24/8/085009} {\bibfield  {journal} {\bibinfo
  {journal} {Journal of Micromechanics and Microengineering}\ }\textbf
  {\bibinfo {volume} {24}},\ \bibinfo {pages} {085009} (\bibinfo {year}
  {2014})}\BibitemShut {NoStop}%
\bibitem [{\citenamefont {Meir}\ \emph {et~al.}(2013)\citenamefont {Meir},
  \citenamefont {Stephanos}, \citenamefont {Geballe},\ and\ \citenamefont
  {Mannhart}}]{meirHighlyefficientThermoelectronicConversion2013}%
  \BibitemOpen
  \bibfield  {author} {\bibinfo {author} {\bibfnamefont {S.}~\bibnamefont
  {Meir}}, \bibinfo {author} {\bibfnamefont {C.}~\bibnamefont {Stephanos}},
  \bibinfo {author} {\bibfnamefont {T.~H.}\ \bibnamefont {Geballe}}, \ and\
  \bibinfo {author} {\bibfnamefont {J.}~\bibnamefont {Mannhart}},\ }\href
  {\doibase 10.1063/1.4817730} {\bibfield  {journal} {\bibinfo  {journal}
  {Journal of Renewable and Sustainable Energy}\ }\textbf {\bibinfo {volume}
  {5}},\ \bibinfo {pages} {043127} (\bibinfo {year} {2013})}\BibitemShut
  {NoStop}%
\bibitem [{\citenamefont {Wanke}\ \emph {et~al.}(2016)\citenamefont {Wanke},
  \citenamefont {Hassink}, \citenamefont {Stephanos}, \citenamefont {Rastegar},
  \citenamefont {Braun},\ and\ \citenamefont
  {Mannhart}}]{wankeMagneticfieldfreeThermoelectronicPower2016}%
  \BibitemOpen
  \bibfield  {author} {\bibinfo {author} {\bibfnamefont {R.}~\bibnamefont
  {Wanke}}, \bibinfo {author} {\bibfnamefont {G.~W.~J.}\ \bibnamefont
  {Hassink}}, \bibinfo {author} {\bibfnamefont {C.}~\bibnamefont {Stephanos}},
  \bibinfo {author} {\bibfnamefont {I.}~\bibnamefont {Rastegar}}, \bibinfo
  {author} {\bibfnamefont {W.}~\bibnamefont {Braun}}, \ and\ \bibinfo {author}
  {\bibfnamefont {J.}~\bibnamefont {Mannhart}},\ }\href {\doibase
  10.1063/1.4955073} {\bibfield  {journal} {\bibinfo  {journal} {Journal of
  Applied Physics}\ }\textbf {\bibinfo {volume} {119}},\ \bibinfo {pages}
  {244507} (\bibinfo {year} {2016})}\BibitemShut {NoStop}%
\bibitem [{\citenamefont {Rasor}(1991)}]{rasorThermionicEnergyConversion1991}%
  \BibitemOpen
  \bibfield  {author} {\bibinfo {author} {\bibfnamefont {N.}~\bibnamefont
  {Rasor}},\ }\href {\doibase 10.1109/27.125041} {\bibfield  {journal}
  {\bibinfo  {journal} {IEEE Transactions on Plasma Science}\ }\textbf
  {\bibinfo {volume} {19}},\ \bibinfo {pages} {1191} (\bibinfo {year}
  {Dec./1991})}\BibitemShut {NoStop}%
\bibitem [{\citenamefont {{Baksht}}\ \emph {et~al.}(1978)\citenamefont
  {{Baksht}}, \citenamefont {{Dyvzhev}}, \citenamefont {{Martsinovskiy}},
  \citenamefont {{Moyzhes}}, \citenamefont {{Dikus}}, \citenamefont {{Sonin}},\
  and\ \citenamefont {{Yuryev}}}]{baksht}%
  \BibitemOpen
  \bibfield  {author} {\bibinfo {author} {\bibfnamefont {F.~G.}\ \bibnamefont
  {{Baksht}}}, \bibinfo {author} {\bibfnamefont {G.~A.}\ \bibnamefont
  {{Dyvzhev}}}, \bibinfo {author} {\bibfnamefont {A.~M.}\ \bibnamefont
  {{Martsinovskiy}}}, \bibinfo {author} {\bibfnamefont {B.~Y.}\ \bibnamefont
  {{Moyzhes}}}, \bibinfo {author} {\bibfnamefont {G.~Y.}\ \bibnamefont
  {{Dikus}}}, \bibinfo {author} {\bibfnamefont {E.~B.}\ \bibnamefont
  {{Sonin}}}, \ and\ \bibinfo {author} {\bibfnamefont {V.~G.}\ \bibnamefont
  {{Yuryev}}},\ }\href@noop {} {\enquote {\bibinfo {title} {{Thermionic
  converters and low-temperature plasma}},}\ }\bibinfo {howpublished} {NASA
  STI/Recon Technical Report N} (\bibinfo {year} {1978})\BibitemShut {NoStop}%
\bibitem [{\citenamefont {Hernqvist}(1963)}]{hernqvistAnalysisArcMode1963}%
  \BibitemOpen
  \bibfield  {author} {\bibinfo {author} {\bibfnamefont {K.}~\bibnamefont
  {Hernqvist}},\ }\href {\doibase 10.1109/PROC.1963.2267} {\bibfield  {journal}
  {\bibinfo  {journal} {Proceedings of the IEEE}\ }\textbf {\bibinfo {volume}
  {51}},\ \bibinfo {pages} {748} (\bibinfo {year} {1963})}\BibitemShut
  {NoStop}%
\bibitem [{\citenamefont {Wilkins}\ and\ \citenamefont
  {Gyftopoulos}(1966)}]{wilkinsThermionicConvertersOperating1966}%
  \BibitemOpen
  \bibfield  {author} {\bibinfo {author} {\bibfnamefont {D.~R.}\ \bibnamefont
  {Wilkins}}\ and\ \bibinfo {author} {\bibfnamefont {E.~P.}\ \bibnamefont
  {Gyftopoulos}},\ }\href {\doibase 10.1063/1.1782146} {\bibfield  {journal}
  {\bibinfo  {journal} {Journal of Applied Physics}\ }\textbf {\bibinfo
  {volume} {37}},\ \bibinfo {pages} {2892} (\bibinfo {year}
  {1966})}\BibitemShut {NoStop}%
\bibitem [{\citenamefont {Oettinger}\ and\ \citenamefont
  {Hussman}(1978)}]{oettingerExperimentsEnhancedMode1978}%
  \BibitemOpen
  \bibfield  {author} {\bibinfo {author} {\bibfnamefont {P.~E.}\ \bibnamefont
  {Oettinger}}\ and\ \bibinfo {author} {\bibfnamefont {F.~N.}\ \bibnamefont
  {Hussman}},\ }\href {\doibase 10.1109/TPS.1978.4317090} {\bibfield  {journal}
  {\bibinfo  {journal} {IEEE Transactions on Plasma Science}\ }\textbf
  {\bibinfo {volume} {6}},\ \bibinfo {pages} {83} (\bibinfo {year}
  {1978})}\BibitemShut {NoStop}%
\bibitem [{\citenamefont {Huffman}\ \emph {et~al.}(1976)\citenamefont
  {Huffman}, \citenamefont {Sommer}, \citenamefont {Balestra}, \citenamefont
  {Briere},\ and\ \citenamefont {Oettinger}}]{huffman1976high}%
  \BibitemOpen
  \bibfield  {author} {\bibinfo {author} {\bibfnamefont {F.}~\bibnamefont
  {Huffman}}, \bibinfo {author} {\bibfnamefont {A.}~\bibnamefont {Sommer}},
  \bibinfo {author} {\bibfnamefont {C.}~\bibnamefont {Balestra}}, \bibinfo
  {author} {\bibfnamefont {D.}~\bibnamefont {Briere}}, \ and\ \bibinfo {author}
  {\bibfnamefont {P.}~\bibnamefont {Oettinger}},\ }\href@noop {} {\  (\bibinfo
  {year} {1976})}\BibitemShut {NoStop}%
\bibitem [{\citenamefont {Zherebtsov}\ and\ \citenamefont
  {Talanova}(1976)}]{Zherebtsov}%
  \BibitemOpen
  \bibfield  {author} {\bibinfo {author} {\bibfnamefont {V.}~\bibnamefont
  {Zherebtsov}}\ and\ \bibinfo {author} {\bibfnamefont {V.}~\bibnamefont
  {Talanova}},\ }\href@noop {} {\bibfield  {journal} {\bibinfo  {journal}
  {Pisma v Zhurnal Tekhnischeskoi Fiziki}\ }\textbf {\bibinfo {volume} {2}},\
  \bibinfo {pages} {124} (\bibinfo {year} {1976})}\BibitemShut {NoStop}%
\bibitem [{\citenamefont {McVey}(1990)}]{mcveyImprovedPulsedIonization1990}%
  \BibitemOpen
  \bibfield  {author} {\bibinfo {author} {\bibfnamefont {J.}~\bibnamefont
  {McVey}},\ }in\ \href {\doibase 10.1109/IECEC.1990.716596} {\emph {\bibinfo
  {booktitle} {Proceedings of the 25th {{Intersociety Energy Conversion
  Engineering Conference}}}}},\ Vol.~\bibinfo {volume} {2}\ (\bibinfo
  {publisher} {{IEEE}},\ \bibinfo {address} {{Reno, Nevada}},\ \bibinfo {year}
  {1990})\ pp.\ \bibinfo {pages} {357--361}\BibitemShut {NoStop}%
\bibitem [{\citenamefont {{Rasor Associates, Inc., Sunnyvale, Calif.
  (USA)}}(1975)}]{rasorassociatesinc.sunnyvalecalif.usaAdvancedThermionicEnergy1975}%
  \BibitemOpen
  \bibfield  {author} {\bibinfo {author} {\bibnamefont {{Rasor Associates,
  Inc., Sunnyvale, Calif. (USA)}}},\ }\href {\doibase 10.2172/7199927}
  {{\selectlanguage {English}\emph {\bibinfo {title} {Advanced Thermionic
  Energy Conversion. {{Progress}} Report, {{September}} 1, 1974--{{August}} 31,
  1975}}}},\ \bibinfo {type} {Tech. Rep.}\ \bibinfo {number} {COO-2263-4,
  7199927}\ (\bibinfo {year} {1975})\BibitemShut {NoStop}%
\bibitem [{\citenamefont {Campanell}\ and\ \citenamefont
  {Umansky}(2017)}]{campanellImprovedUnderstandingHot2017}%
  \BibitemOpen
  \bibfield  {author} {\bibinfo {author} {\bibfnamefont {M.~D.}\ \bibnamefont
  {Campanell}}\ and\ \bibinfo {author} {\bibfnamefont {M.~V.}\ \bibnamefont
  {Umansky}},\ }\href {\doibase 10.1088/1361-6595/aa97a9} {\bibfield  {journal}
  {\bibinfo  {journal} {Plasma Sources Science and Technology}\ }\textbf
  {\bibinfo {volume} {26}},\ \bibinfo {pages} {124002} (\bibinfo {year}
  {2017})}\BibitemShut {NoStop}%
\bibitem [{\citenamefont {Wolff}\ \emph {et~al.}(1990)\citenamefont {Wolff},
  \citenamefont {Veltkamp}, \citenamefont {Schoonen},\ and\ \citenamefont
  {Hendriksen}}]{eindhovenConference1989}%
  \BibitemOpen
  \bibinfo {editor} {\bibfnamefont {L.}~\bibnamefont {Wolff}}, \bibinfo
  {editor} {\bibfnamefont {W.}~\bibnamefont {Veltkamp}}, \bibinfo {editor}
  {\bibfnamefont {J.}~\bibnamefont {Schoonen}}, \ and\ \bibinfo {editor}
  {\bibfnamefont {H.}~\bibnamefont {Hendriksen}},\ eds.,\ \href@noop {}
  {{\selectlanguage {English}\emph {\bibinfo {title} {Thermionic energy
  conversion : specialist conference Eindhoven, The Netherlands October 11-12,
  1989 : proceedings}}}}\ (\bibinfo  {publisher} {Eindhoven University of
  Technology},\ \bibinfo {year} {1990})\ pp.\ \bibinfo {pages}
  {71--95}\BibitemShut {NoStop}%
\bibitem [{\citenamefont {Warner}\ and\ \citenamefont
  {Hansen}(1967)}]{warnerTransportEffectsElectron1967}%
  \BibitemOpen
  \bibfield  {author} {\bibinfo {author} {\bibfnamefont {C.}~\bibnamefont
  {Warner}}\ and\ \bibinfo {author} {\bibfnamefont {L.~K.}\ \bibnamefont
  {Hansen}},\ }\href {\doibase 10.1063/1.1709363} {\bibfield  {journal}
  {\bibinfo  {journal} {Journal of Applied Physics}\ }\textbf {\bibinfo
  {volume} {38}},\ \bibinfo {pages} {491} (\bibinfo {year} {1967})}\BibitemShut
  {NoStop}%
\bibitem [{war()}]{warp}%
  \BibitemOpen
  \href@noop {} {}\bibinfo {note} {See \url{http://warp.lbl.gov/} for code
  details}\BibitemShut {NoStop}%
\bibitem [{\citenamefont {Friedman}\ \emph {et~al.}(1992)\citenamefont
  {Friedman}, \citenamefont {Grote},\ and\ \citenamefont
  {Haber}}]{friedmanThreeDimensionalParticle1992}%
  \BibitemOpen
  \bibfield  {author} {\bibinfo {author} {\bibfnamefont {A.}~\bibnamefont
  {Friedman}}, \bibinfo {author} {\bibfnamefont {D.~P.}\ \bibnamefont {Grote}},
  \ and\ \bibinfo {author} {\bibfnamefont {I.}~\bibnamefont {Haber}},\ }\href
  {\doibase 10.1063/1.860024} {\bibfield  {journal} {\bibinfo  {journal}
  {Physics of Fluids B: Plasma Physics}\ }\textbf {\bibinfo {volume} {4}},\
  \bibinfo {pages} {2203} (\bibinfo {year} {1992})}\BibitemShut {NoStop}%
\bibitem [{\citenamefont {Grote}(2005)}]{groteWARPCodeModeling2005}%
  \BibitemOpen
  \bibfield  {author} {\bibinfo {author} {\bibfnamefont {D.~P.}\ \bibnamefont
  {Grote}},\ }in\ \href {\doibase 10.1063/1.1893366} {{\selectlanguage
  {English}\emph {\bibinfo {booktitle} {{{AIP Conference Proceedings}}}}}},\
  Vol.\ \bibinfo {volume} {749}\ (\bibinfo  {publisher} {{AIP}},\ \bibinfo
  {address} {{Berkeley, California (USA)}},\ \bibinfo {year} {2005})\ pp.\
  \bibinfo {pages} {55--58}\BibitemShut {NoStop}%
\bibitem [{\citenamefont {Birdsall}(1985)}]{birdsall1985plasma}%
  \BibitemOpen
  \bibfield  {author} {\bibinfo {author} {\bibfnamefont {C.}~\bibnamefont
  {Birdsall}},\ }\href@noop {} {\emph {\bibinfo {title} {Plasma physics via
  computer simulation}}}\ (\bibinfo  {publisher} {McGraw-Hill},\ \bibinfo
  {address} {New York},\ \bibinfo {year} {1985})\BibitemShut {NoStop}%
\bibitem [{\citenamefont {Li}(2005)}]{li05}%
  \BibitemOpen
  \bibfield  {author} {\bibinfo {author} {\bibfnamefont {X.~S.}\ \bibnamefont
  {Li}},\ }\href@noop {} {\bibfield  {journal} {\bibinfo  {journal} {ACM Trans.
  Math. Software}\ }\textbf {\bibinfo {volume} {31}},\ \bibinfo {pages} {302}
  (\bibinfo {year} {2005})}\BibitemShut {NoStop}%
\bibitem [{\citenamefont
  {Birdsall}(1991)}]{birdsallParticleincellChargedparticleSimulations1991}%
  \BibitemOpen
  \bibfield  {author} {\bibinfo {author} {\bibfnamefont {C.}~\bibnamefont
  {Birdsall}},\ }\href {\doibase 10.1109/27.106800} {\bibfield  {journal}
  {\bibinfo  {journal} {IEEE Transactions on Plasma Science}\ }\textbf
  {\bibinfo {volume} {19}},\ \bibinfo {pages} {65} (\bibinfo {year}
  {1991})}\BibitemShut {NoStop}%
\bibitem [{\citenamefont {Turner}\ \emph {et~al.}(2013)\citenamefont {Turner},
  \citenamefont {Derzsi}, \citenamefont {Donk{\'o}}, \citenamefont {Eremin},
  \citenamefont {Kelly}, \citenamefont {Lafleur},\ and\ \citenamefont
  {Mussenbrock}}]{turnerSimulationBenchmarksLowpressure2013}%
  \BibitemOpen
  \bibfield  {author} {\bibinfo {author} {\bibfnamefont {M.~M.}\ \bibnamefont
  {Turner}}, \bibinfo {author} {\bibfnamefont {A.}~\bibnamefont {Derzsi}},
  \bibinfo {author} {\bibfnamefont {Z.}~\bibnamefont {Donk{\'o}}}, \bibinfo
  {author} {\bibfnamefont {D.}~\bibnamefont {Eremin}}, \bibinfo {author}
  {\bibfnamefont {S.~J.}\ \bibnamefont {Kelly}}, \bibinfo {author}
  {\bibfnamefont {T.}~\bibnamefont {Lafleur}}, \ and\ \bibinfo {author}
  {\bibfnamefont {T.}~\bibnamefont {Mussenbrock}},\ }\href {\doibase
  10.1063/1.4775084} {\bibfield  {journal} {\bibinfo  {journal} {Physics of
  Plasmas}\ }\textbf {\bibinfo {volume} {20}},\ \bibinfo {pages} {013507}
  (\bibinfo {year} {2013})}\BibitemShut {NoStop}%
\bibitem [{\citenamefont
  {Phelps}(1994)}]{phelpsApplicationScatteringCross1994}%
  \BibitemOpen
  \bibfield  {author} {\bibinfo {author} {\bibfnamefont {A.~V.}\ \bibnamefont
  {Phelps}},\ }\href {\doibase 10.1063/1.357820} {\bibfield  {journal}
  {\bibinfo  {journal} {Journal of Applied Physics}\ }\textbf {\bibinfo
  {volume} {76}},\ \bibinfo {pages} {747} (\bibinfo {year} {1994})}\BibitemShut
  {NoStop}%
\bibitem [{\citenamefont {Lawless}\ and\ \citenamefont
  {Lam}(1986)}]{lawlessAnalyticalModelThermionic1986}%
  \BibitemOpen
  \bibfield  {author} {\bibinfo {author} {\bibfnamefont {J.~L.}\ \bibnamefont
  {Lawless}}\ and\ \bibinfo {author} {\bibfnamefont {S.~H.}\ \bibnamefont
  {Lam}},\ }\href {\doibase 10.1063/1.336416} {\bibfield  {journal} {\bibinfo
  {journal} {Journal of Applied Physics}\ }\textbf {\bibinfo {volume} {59}},\
  \bibinfo {pages} {1875} (\bibinfo {year} {1986})}\BibitemShut {NoStop}%
\bibitem [{\citenamefont {Campanell}(2018)}]{PhysRevE.97.043207}%
  \BibitemOpen
  \bibfield  {author} {\bibinfo {author} {\bibfnamefont {M.~D.}\ \bibnamefont
  {Campanell}},\ }\href {\doibase 10.1103/PhysRevE.97.043207} {\bibfield
  {journal} {\bibinfo  {journal} {Phys. Rev. E}\ }\textbf {\bibinfo {volume}
  {97}},\ \bibinfo {pages} {043207} (\bibinfo {year} {2018})}\BibitemShut
  {NoStop}%
\bibitem [{\citenamefont {Greiner}\ \emph {et~al.}(1995)\citenamefont
  {Greiner}, \citenamefont {Klinger},\ and\ \citenamefont
  {Piel}}]{greinerNonlinearDynamicalBehavior1995}%
  \BibitemOpen
  \bibfield  {author} {\bibinfo {author} {\bibfnamefont {F.}~\bibnamefont
  {Greiner}}, \bibinfo {author} {\bibfnamefont {T.}~\bibnamefont {Klinger}}, \
  and\ \bibinfo {author} {\bibfnamefont {A.}~\bibnamefont {Piel}},\ }\href
  {\doibase 10.1063/1.871335} {\bibfield  {journal} {\bibinfo  {journal}
  {Physics of Plasmas}\ }\textbf {\bibinfo {volume} {2}},\ \bibinfo {pages}
  {1810} (\bibinfo {year} {1995})}\BibitemShut {NoStop}%
\bibitem [{\citenamefont {Levko}(2015)}]{levkoInfluenceAnodeProcesses2015}%
  \BibitemOpen
  \bibfield  {author} {\bibinfo {author} {\bibfnamefont {D.}~\bibnamefont
  {Levko}},\ }\href {\doibase 10.1063/1.4923463} {\bibfield  {journal}
  {\bibinfo  {journal} {Physics of Plasmas}\ }\textbf {\bibinfo {volume}
  {22}},\ \bibinfo {pages} {073501} (\bibinfo {year} {2015})}\BibitemShut
  {NoStop}%
\bibitem [{\citenamefont {Levush}\ \emph {et~al.}(2000)\citenamefont {Levush},
  \citenamefont {Lau},\ and\ \citenamefont
  {Jensen}}]{levushMigrationEscapeBarium2000}%
  \BibitemOpen
  \bibfield  {author} {\bibinfo {author} {\bibfnamefont {B.}~\bibnamefont
  {Levush}}, \bibinfo {author} {\bibfnamefont {Y.}~\bibnamefont {Lau}}, \ and\
  \bibinfo {author} {\bibfnamefont {K.}~\bibnamefont {Jensen}},\ }\href
  {\doibase 10.1109/27.887721} {\bibfield  {journal} {\bibinfo  {journal} {IEEE
  Transactions on Plasma Science}\ }\textbf {\bibinfo {volume} {28}},\ \bibinfo
  {pages} {772} (\bibinfo {year} {2000})}\BibitemShut {NoStop}%
\bibitem [{\citenamefont
  {Desplat}(1997)}]{desplatRecentDevelopmentsOxygenated1997}%
  \BibitemOpen
  \bibfield  {author} {\bibinfo {author} {\bibfnamefont {J.-L.}\ \bibnamefont
  {Desplat}},\ }in\ \href {\doibase 10.1016/B978-044482548-3/50105-0}
  {{\selectlanguage {English}\emph {\bibinfo {booktitle} {Functionally {{Graded
  Materials}} 1996}}}}\ (\bibinfo  {publisher} {{Elsevier}},\ \bibinfo {year}
  {1997})\ pp.\ \bibinfo {pages} {639--646}\BibitemShut {NoStop}%
\bibitem [{\citenamefont {Verboncoeur}\ \emph {et~al.}(1995)\citenamefont
  {Verboncoeur}, \citenamefont {Langdon},\ and\ \citenamefont
  {Gladd}}]{verboncoeurObjectorientedElectromagneticPIC1995}%
  \BibitemOpen
  \bibfield  {author} {\bibinfo {author} {\bibfnamefont {J.}~\bibnamefont
  {Verboncoeur}}, \bibinfo {author} {\bibfnamefont {A.}~\bibnamefont
  {Langdon}}, \ and\ \bibinfo {author} {\bibfnamefont {N.}~\bibnamefont
  {Gladd}},\ }\href {\doibase 10.1016/0010-4655(94)00173-Y} {\bibfield
  {journal} {\bibinfo  {journal} {Computer Physics Communications}\ }\textbf
  {\bibinfo {volume} {87}},\ \bibinfo {pages} {199} (\bibinfo {year}
  {1995})}\BibitemShut {NoStop}%
\end{thebibliography}%

\end{document}